\documentclass[preprint]{emulateapj}
\usepackage{hyperref}
\usepackage[usenames,dvipsnames]{color}
\hypersetup{colorlinks,citecolor=Blue,linkcolor=Red,urlcolor=Blue}
\usepackage[titletoc]{appendix}
\usepackage{amsmath}

\usepackage{graphicx}
\usepackage[T1]{fontenc}
\usepackage[utf8]{inputenc}

\begin{document} 

\title{The Primordial Solar wind as a Sculptor of Terrestrial Planet Formation}

\author{Christopher Spalding$^{1}$} 
%\affil{$^1$Division of Geological and Planetary Sciences\\
%California Institute of Technology, Pasadena, CA 91125} 
\affil{$^1$Department of Astronomy, Yale University, New Haven, CT 06511} 

%\begin{document}
%\maketitle
%\small
\begin{abstract}
Our Solar System is almost entirely devoid of material interior to Mercury's orbit, in sharp contrast to the multiple Earth masses of material commonly residing within the analogous region of extrasolar planetary systems. Recent work has suggested that Jupiter's orbital migration early in the Solar System's history fragmented primordial planetary material within the inner Solar System. However, the reason for the absence of subsequent planet formation within 0.4\,AU remains unsolved. Here, we show that the up debris interior to Mercury's current orbit is susceptible to outward migration driven by the early Solar wind, enhanced by the Sun's primordial rapid rotation and strong magnetic field. The ram pressure arising from azimuthal motion of the Solar wind plasma transported $\sim100$\,m-sized objects and smaller from 0.1\,AU out to the terrestrial planet-forming zone within the suspected $\sim30-50$\,Myr timespan of the Earth's formation. The mass of material within this size class typically exceeds Mercury, and can rival that of Earth. Consequently, the present-day region of terrestrial planets and the asteroid belt has been supplied by a large mass of material from the innermost, hot Solar System, providing a potential explanation for the evidence of high-temperature alteration within some asteroids and the high iron content of Mercury.
\end{abstract}

%\thispagestyle{firststyle}
%\ifthenelse{\boolean{shortarticle}}{\ifthenelse{\boolean{singlecolumn}}{\abscontentformatted}{\abscontent}}{}

\section{Introduction}

A recent influx of planets discovered orbiting other stars has shed light upon the Solar System's peculiarities. Specifically, a hallmark of many extrasolar planetary systems is their proximity to the host star. Commonly, planets larger than Earth occupy orbits significantly closer-in than Mercury, contrasting sharply with the emptiness of our own inner Solar System \citep{Chiang2013,Coughlin2016}. 

An important clue to this puzzling disparity is that close-in ``super-Earths" around other stars often possess extensive atmospheres, and thus necessarily formed before the dissipation of their natal gas disks; within about 3\,Myr \citep{Haisch2001}. In contrast, isotopic evidence reveals that the Earth formed in roughly $\sim30-50$\,Myr \citep{Kleine2009,Yu2011}. Somehow, terrestrial planet formation in our Solar System was inhibited while the gas persisted, requiring a slower, gas-free mode of planet-formation to assemble the innermost 3 planets \citep{Morbidelli2012b}. 

A widely-cited history of our Solar System holds that Jupiter underwent gas-driven inward migration to around 1.5\,AU, before subsequently migrating outwards due to interactions with Saturn \citep{Hansen2009,Walsh2011}. This scenario has been augmented by the suggestion that Jupiter's trek swept up primordial planetesimals or planets into mean motion resonances, exciting their eccentricities and causing destructive collisions \citep{Batygin2015b}. Accordingly, our Solar System's lack of extensive planet formation interior to $\sim1$\,AU, prior to disk-dispersal, emerges as a consequence of Jupiter's influence. 

Jupiter's migration addresses the absence of super-Earths in the inner Solar System, but does not explain the lack of subsequent planet formation interior to $\sim0.3$\,AU. Jupiter would have initiated a collisional cascade, resulting in $\sim20$ Earth masses of debris, occupying size classes $\lesssim$\,100\,km \citep{Batygin2015b}, with significant mass occupying 100\,m and smaller. Such objects aerodynamically lose angular momentum when embedded in a gas disk \citep{Weidenschilling1977}, however, the gas only extended to roughly 0.05-0.1\,AU from the Sun's surface \citep{Armitage2011}, suggesting that aerodynamic drag cannot entirely remove this debris. 

 The above picture ignores the Solar wind, which today consists of a stream of ionized plasma, moving radially outwards into space at several hundred kms$^{-1}$ \citep{Phillips1995}. As orbiting debris encounters this plasma today, it faces a headwind. However, a critical distinction in the early Solar System is that the young Sun would have possessed a significantly enhanced magnetic field and faster rotation rate \citep{Bouvier2013,Folsom2017,Fion2018}. The resulting coupling between the plasma and magnetosphere generated a large azimuthal component to the early Solar wind--a tailwind. This tailwind lends angular momentum to the orbiting debris, causing the orbits to grow. 

In this work, we propose that the Solar wind drove outward migration of debris interior to 0.4\,AU, both inhibiting planet formation there, and enriching the terrestrial planets with material of high-temperature origin. 

\begin{figure}[h!]
\centering
\includegraphics[trim=0cm 0cm 0cm 0cm, clip=true,width=1\columnwidth]{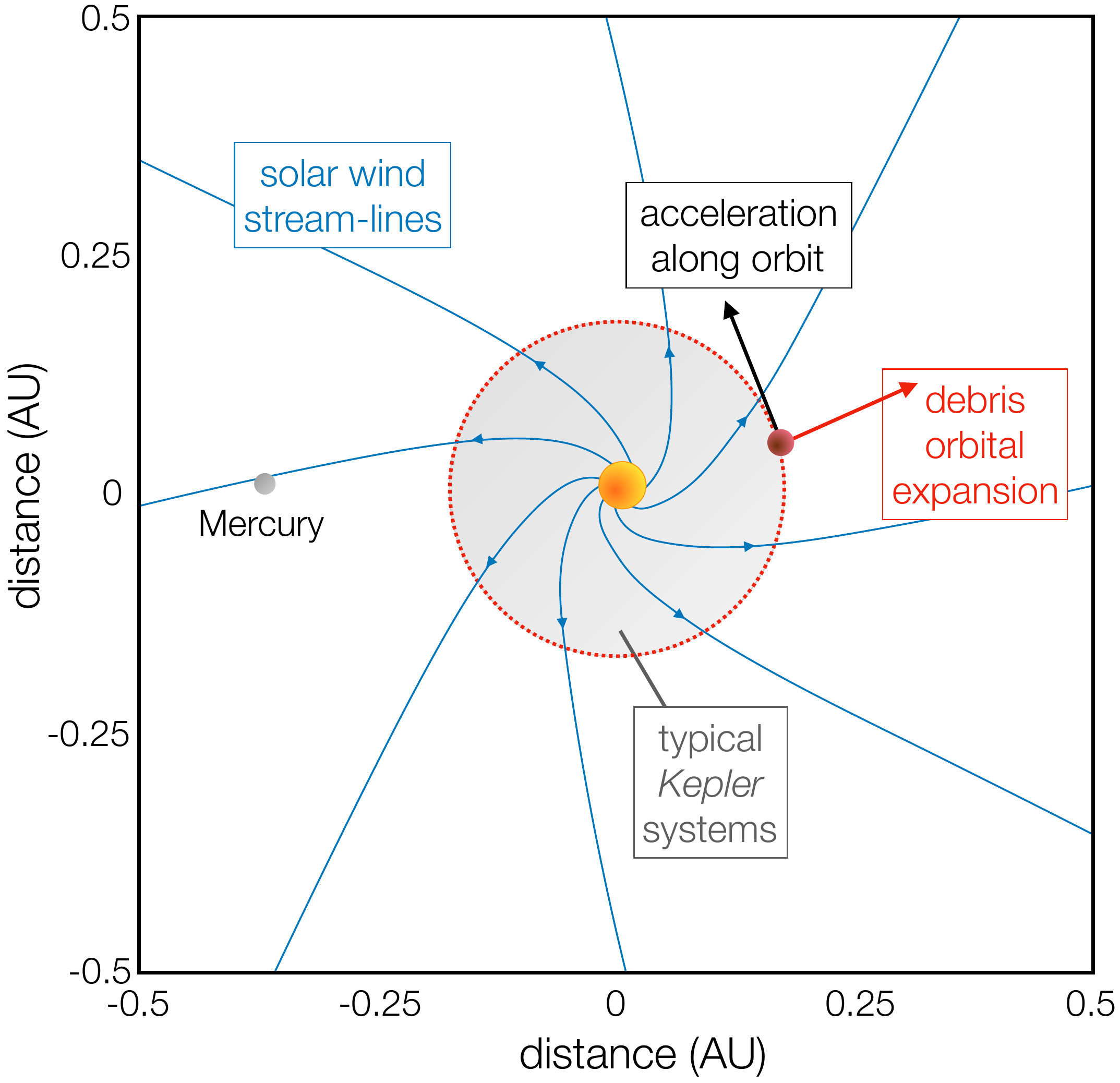}
\caption{A schematic of the Solar wind plasma trajectory. The blue spirals represent streamlines of wind plasma at an age of 30\,Myr. Debris following the red orbit experiences a tailwind, expanding its orbit beyond that typical of extrasolar planets (gray). Mercury's orbital distance is shown for reference.}
\label{Schematic}
\end{figure}

Section 2 describes a simple, axisymmetric model representing the azimuthal velocity of the wind plasma (\citealt{Weber1967,Lovelace2008}; shown schematically in Figure~\ref{Schematic}). The magnetic field, spin rate and mass-loss rates appropriate to the early Sun are obtained from empirical measurements and model calculations of Sun-like stars \citep{Folsom2017,Fion2018}. We then discuss the implications for the Solar wind for the terrestrial planets' formation.

\section{Methods}
\subsection{Solar wind Properties}

Today, the Sun loses mass through the Solar wind at a rate of $2\times10^{-14}\,M_\odot\,$yr$^{-1}$ \citep{Phillips1995}, and consists of  plasma flowing radially outward. Early models reproducing this radial velocity are attributed to \citet{Parker1965}, and ignore significant azimuthal components to the wind plasma. Whereas the neglect of azimuthal motion is appropriate for today's Solar System, young stars possess strong magnetic fields and faster rotations, lending the plasma a significant azimuthal velocity. 

A self-consistent calculation of azimuthal and radial velocities, $v_\phi$ and $v_r$, as a function of heliocentric distance $r$ has been computed elsewhere \citep{Weber1967,Hartmann1982}. Following this previous work, we assume that the central star possesses a radial magnetic field of the form
\begin{align}
B_r(r,t)=B_\star(t)\bigg(\frac{r}{R_\star}\bigg)^{-2},
\end{align}
where $t$ is time and $R_\star=2R_\odot$ is the radius of the stellar surface, approximated as twice the current Solar radius \citep{Shu1987,Armitage1996}, and $B_\star(t)$ is the magnetic field strength at the surface of the star. 

The modeled wind is azimuthally symmetric and possesses a mass-loss rate $\dot{M}$, yielding a plasma density of
\begin{align}\label{density}
\rho_{sw}=\frac{\dot{M}}{4\pi\,v_r\,r^2}.
\end{align}
Stellar rotation, with period $P_\star(t)$, attempts to translate magnetospheric field lines relative to the plasma, inducing currents that generate azimuthal magnetic fields, which are the ultimate sources of the azimuthal velocity. At very small $r$, strong coupling drives near-corotation between the plasma and the star. Qualitatively, this region lies inside the Alfv\'en radius, approximated by \citep{Lovelace2008}
\begin{align}
r_A(t)\approx\bigg(\frac{3}{8}\bigg)^{1/2}v_M(t)P_\star(t)
\end{align}
where we define the magnetic velocity \citep{Michel1969} as
\begin{align}
v_M(t)\equiv\Bigg(\frac{16\pi^3B_\star(t)^2R_\star^4 }{\mu_0\dot{M}(t)P_\star(t)^2}\Bigg)^{1/3},
\end{align}
$\mu_0$ is the vacuum permeability, and explicit time-dependence is retained.

 Rather than solving stellar wind models from first principles, we adopt the approximate functional forms presented in \citet{Lovelace2008}:
 \begin{align}
 v_r&=\frac{2}{3}v_M\Bigg(\frac{1.8r/r_A}{1+0.8r/r_A}\Bigg)\nonumber\\
 v_\phi&=\frac{2}{3}v_M\Bigg(\frac{1.8r/r_A}{(1+1.5r/r_A)^2}\Bigg).
 \end{align}
 These expressions are strictly valid only when $r\lesssim\,r_A$, however, at larger radii their accuracy suffices to the degree required here, given the uncertainties inherent to the relevant stellar parameters. 
 
 Debris in orbit around the young star will experience a tailwind if $v_\phi>v_K=\sqrt{G M_\star/r}$, where the Keplerian orbital velocity $v_K$ is written in terms of Newton's gravitational constant $G$ and stellar mass $M_\star$, which we set to equal one Solar mass ($M_\star=M_\odot$). We now present a calculation of the force experienced by particles in the wind.

\subsection{Solar wind ram pressure}

We consider a spherical particle of radius $s$ and density $\rho_s=3000$\,kg\,m$^{-3}$, following a circular orbit of semi-major axis $r$ and  velocity $\mathbf{v}_K$. The solar wind exerts a ram pressure of $C_D\rho_{\textrm{sw}} |\mathbf{v}_{\textrm{sw}}-\mathbf{v}_K|^2$, where the solar wind velocity is denoted $\mathbf{v}_{\textrm{sw}}$, and $C_D$ is a drag coefficient. The exact value of $C_D$ varies with composition and grain size \citep{Mukai1982}, but for the precision demanded here we set $C_D=1$.  

After multiplying by the particle's wind-facing surface area $\pi\,s^2$, the azimuthal component of the incident force $F_\phi$ takes the form
\begin{align}
F_\phi&=C_D\pi s^2\rho_{\textrm{sw}}|\mathbf{v}_{\textrm{sw}}-\mathbf{v}_p|\big(v_\phi-\Omega\,r\big)\nonumber\\
&=C_D\pi\,s^2 \rho_{\textrm{sw}}\sqrt{v_r^2+(v_\phi-\Omega\,r\big)^2}\big(v_\phi-\Omega\,r\big),
\end{align} 
where $\Omega$ is the orbital angular velocity.
 
Combining with equation~\ref{density}, the angular momentum loss rate (torque) upon the planetesimal is
\begin{align}
\dot{L}&=F_\phi\,r\nonumber\\
&=rC_D\bigg(\frac{s^2}{4r^2}\bigg)\bigg(\frac{\dot{M}}{v_r}\bigg)\sqrt{v_r^2+(v_\phi-\Omega\,r\big)^2}\big(v_\phi-\Omega\,r\big).
\end{align}
The angular momentum of a particle of mass $m$ is $m\sqrt{GM_\star\,r}$, such that the orbital drift timescale reads
\begin{align}
&\tau_r\equiv\frac{r}{\dot{r}}=\frac{L}{2\dot{L}}\nonumber\\
&=\frac{8\pi\rho_s\Omega\,v_rsr^3}{3C_D\dot{M}\bigg[\sqrt{v_r^2+(v_\phi-\Omega\,r\big)^2}\big(v_\phi-\Omega\,r\big)\bigg]}.
\end{align}
Note that if $v_\phi>\Omega\,r$, the particle's orbit will expand outwards and vice versa.

\subsection{Early stellar properties}

The magnitude of stellar winds from Sun-like stars is difficult to reliably measure \citep{Gaidos2000,Wood2002,Wood2005,Wood2014,Bouvier2014,Vidotto2014}. Using Ly$\alpha$ signatures, produced when stellar winds interact with the interstellar medium, the loss rates of several Sun-like stars have been measured \citep{Wood2002,Wood2005,Wood2014}. These measurements are sensitive to assumptions associated with modeling ISM interactions, such as the generation of a bow shock, but suggest that winds of Sun-like stars generally decrease over time, following an approximate relationship of $\dot{M}\propto t^{-2}$ \citep{Wood2002}. However, this expression appears not to apply younger than $\sim700$\,Myr, when a different, and less predictable relationship holds. During this early phase, observations are severely lacking, and so we adopt the model results of \citet{Fion2018}, which are consistent with the measurements of \citet{Wood2005} at later epochs. These relationships take the following form:
\begin{align}
\dot{M}(t)&=5\times10^{-10}\Bigg(\frac{t}{\textrm{Myr}}\Bigg)^{-3/4}M_\odot\textrm{yr}^{-1}\nonumber\\
B_\star(t)&=500\Bigg(\frac{t}{\textrm{Myr}}\Bigg)^{-1/2}\textrm{Gauss}\nonumber\\
P_\star(t)&=\frac{2}{3}\Bigg(\frac{t}{\textrm{Myr}}\Bigg)^{1/2}\textrm{days}.
\end{align}

The equations above are likely to vary considerably from system to system, however, for the sake of definiteness and analytical tractability, we utilize the above forms throughout. Below, we will analyze the system as time progresses, during which, the values of the physical parameters $v_M$ and $r_A$ change, effectively exploring a range of parameter regimes. 

\subsection{Comparison to Poynting-Robertson drag}

In addition to the Solar wind, the Sun's photon radiation exerts an azimuthal force upon orbiting material, known as Poynting-Robertson drag, of magnitude \citep{Gustafson1994}
\begin{align}
F_{\textrm{PR}}=\frac{L_\star\,s^2}{4r^2}\frac{v_K}{c^2},
\end{align}
where we introduce the speed of light $c$ and stellar luminosity $L_\star$.

Assuming $v_\phi=0$ and $v_r\gg v_K$, the ratio of Poynting-Robertson to solar wind drag today is
\begin{align}
\frac{F_{\phi,0}}{F_{\textrm{PR}}}&=\frac{\dot{M}}{L_\star/c^2}\frac{C_D}{C_p}\nonumber\\
&\approx0.3\bigg(\frac{\dot{M}}{\dot{M}_\odot}\bigg)\bigg(\frac{L_\star}{L_\odot}\bigg)^{-1}\bigg(\frac{C_D}{C_p}\bigg),
\end{align}
using as nominal values $\dot{M}_\odot=2\times10^{-14}M_\odot$yr$^{-1}$ and $L_\odot=3.8\times10^{26}$W \citep{Wood2014,Genova2018}. Accordingly, Poynting-Robertson drag exceeds solar wind pressure in the current Solar System. However, as we show below, the enhanced azimuthal motion of the solar wind during early epochs causes solar wind drag to dominate during close-in planet formation. 

\section{Results}
\begin{figure}[h!]
\centering
\includegraphics[trim=0cm 0cm 0cm 0cm, clip=true,width=1\columnwidth]{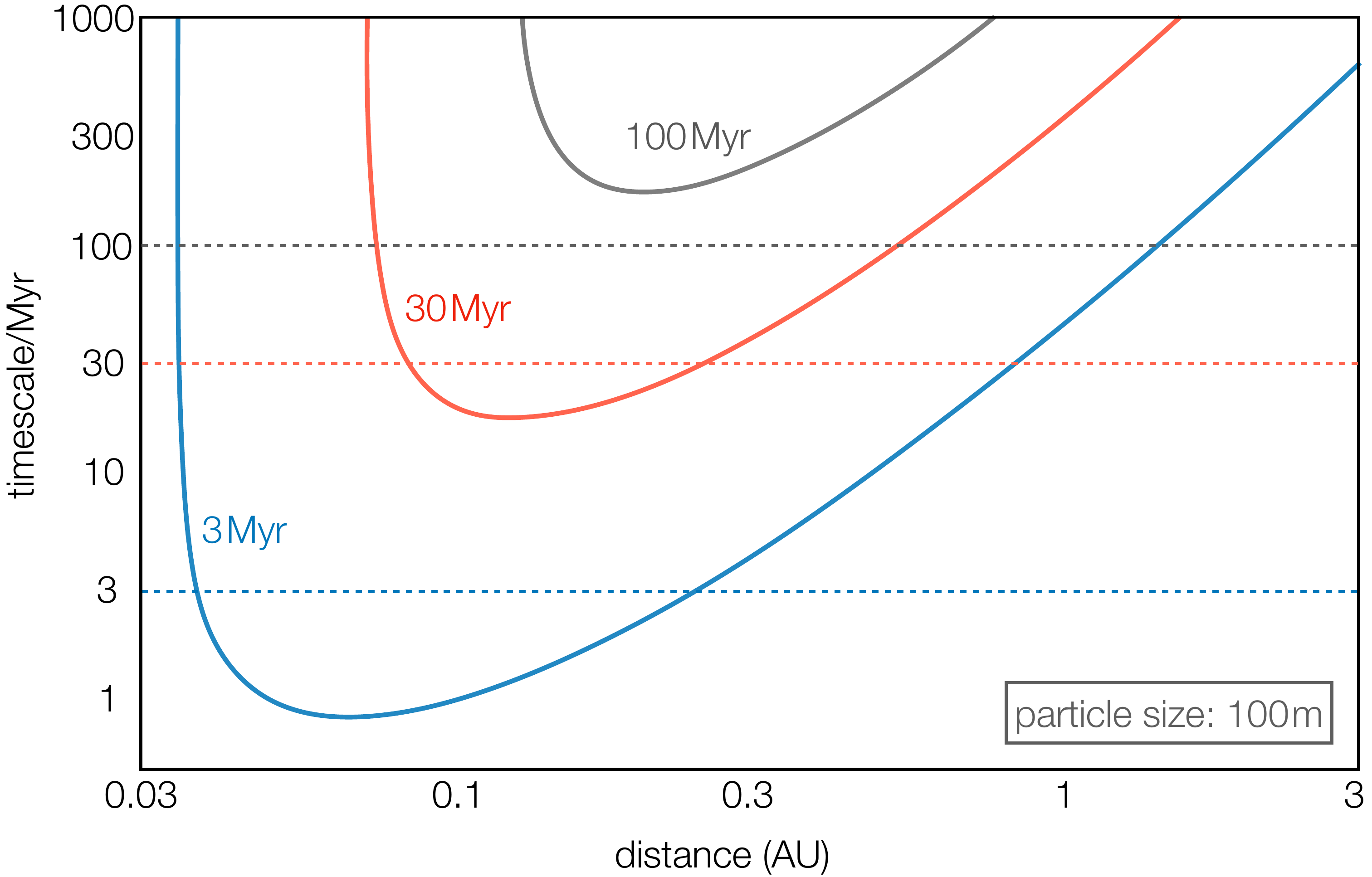}
\caption{Solar wind-induced outward drift timescale for particles of size 100\,m at 3 epochs: 3\,Myr, 30\,Myr and 100\,Myr. Horizontal lines denote these 3 times in order to compare the drift timescales to the system ages (since disk-dispersal). Notice that drift timescales remain shorter than system ages beyond $\sim30\,$Myr.}
\label{f2}
\end{figure}

Using the parameters described above, and neglecting Poynting-Robertson drag for now, we present the timescale over which particle orbits grow in Figure~\ref{f2}. We illustrate timescales relevant for 100\,m objects, but the migration time scales linearly with particle size. Results are presented at 3 epochs; 3\,Myr, 30\,Myr and 100\,Myr subsequent to disk-dispersal (where the disk is assumed to disperse at 3\,Myr of age). Drift times increase as the star's rotation slows, its magnetic field weakens, and its mass-loss decays. 

On Figure~\ref{f2}, we plot horizontal lines corresponding to the ages of the 3 epochs. Qualitatively, wherever the drift timescale is less than the age, significant outward migration is expected. As time progresses, the drift timescale increases faster than the system age, such that at some critical time, no orbital regions undergo migration over timescales shorter than the time since disk dispersal; written quantitatively as $\tau_r(t)=t-\tau_{\textrm{disk}}$. For example, by 100\,Myr, 100\,m particles migrate too slowly to experience any significant drift over their lifetimes. 

\begin{figure}
\centering
\includegraphics[trim=0cm 0cm 0cm 0cm, clip=true,width=1\columnwidth]{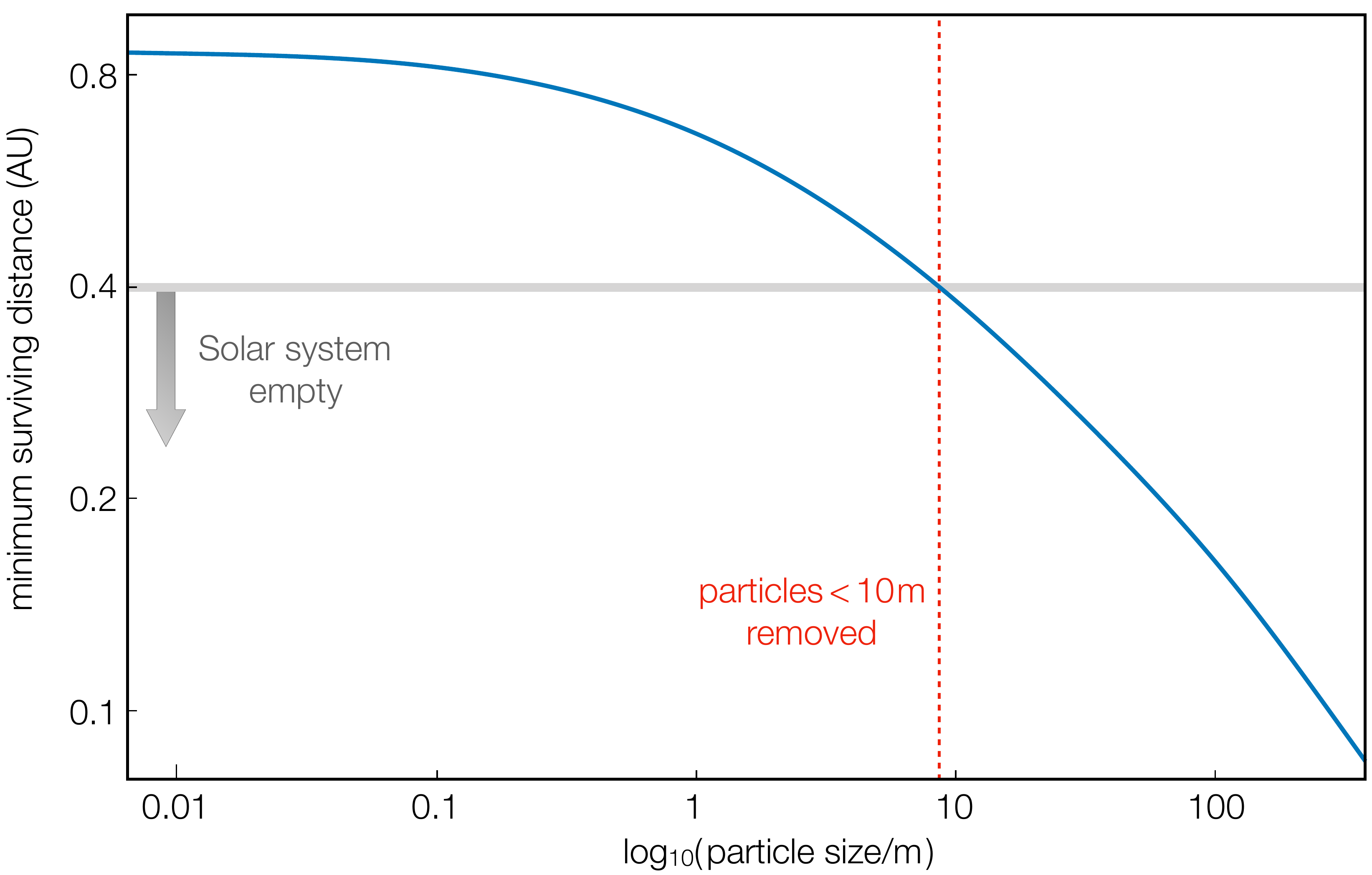}
\caption{Minimum orbital distance at which particles survive outward drift, as a function of particle size. These distances are computing by finding the time since disk dissipation that the drift time is longer than system age (Figure~\ref{f2}) and computing the corresponding minimum orbital distance. Particles below 10\,m inside are entirely removed form inside Mercury's orbit.}
\label{f3}
\end{figure}

In Figure~\ref{f3}, we plot, as a function of particle size, the orbital distance at which the drift time equals the time since disk dispersal. This may be thought of as the minimum distance at which particles of a given size may survive the influence of the Solar wind. This figure suggests that particles smaller than $\sim10$\,m are unable to survive interior to Mercury's orbit. However, this constitutes a conservative estimate, given that most migration is likely to occur earlier than the point at which drift time equals system age. In what follows, we show that 100\,m objects are likely to be efficiently removed under a more detailed treatment. Nevertheless, Figure~\ref{f3} serves as a useful order of magnitude estimate for the lower-limit on surviving particle sizes interior to a given orbital distance.

The timescales presented in Figure~\ref{f2} relate to unmagnetized particles. However, once planetesimals coalesce to form planetary bodies, they generate their own strong magnetic fields, augmenting star-planet torques. It was shown by \citet{Lovelace2008} that a Jupiter-like object with magnetic field 100\,Gauss, orbiting near the co-rotation radius of a Sun-like star will undergo radial migration over a timescale of $T_{\textrm{w}}\approx20\,$Myr$(r/0.06\,$AU)$^{13/6}$, where 1\,kG was assumed for the stellar field strength. This regime is beyond the scope of our work here, but it is important to briefly note that the central star's magnetic properties may continue sculpting planetary systems subsequent to their initial formation.

\subsection{Simulations}
The static, timescale-driven discussion presented in the previous sections suggest that significant drift may be induced in particles of $\sim100$\,metres in size, from the region interior to Mercury's orbit to the terrestrial planet-forming region. We now present simple, numerical simulations to demonstrate the outward drift of these particles. For completeness, our simulations included the effect of both Poynting-Robertson drag and the Solar wind. We numerically solve 
\begin{align}
\frac{dr}{dt}=\frac{r}{\tau_r}-\frac{r}{\tau_{PR}},
\end{align}
where $\tau_{\textrm{PR}}$ is the timescale of semi-major axis decay associated with Poynting-Robertson drag, $F_{PR}$ as defined above, using the current solar luminosity. As before, simulations begin at 3\,Myr into the Solar System's history, owing to a typical disk dispersal time of 3\,Myr \citep{Haisch2001}. 

In order to deduce the trajectories of orbiting material under the influence of an early solar wind, we simulate 3 particle size classes; 10\,m, 100\,m and 1000\,m, each beginning at 3 different orbital distances; 0.1\,AU, 0.4\,AU and 1\,AU. This range of distances allows us to anaylze the fate of particles beginning interior to Mercury, close to Mercury's present location, and those close to Earth's present location, respectively. All other parameters are kept the same as previous sections.

\begin{figure}
\centering
\includegraphics[trim=0cm 0cm 0cm 0cm, clip=true,width=1\columnwidth]{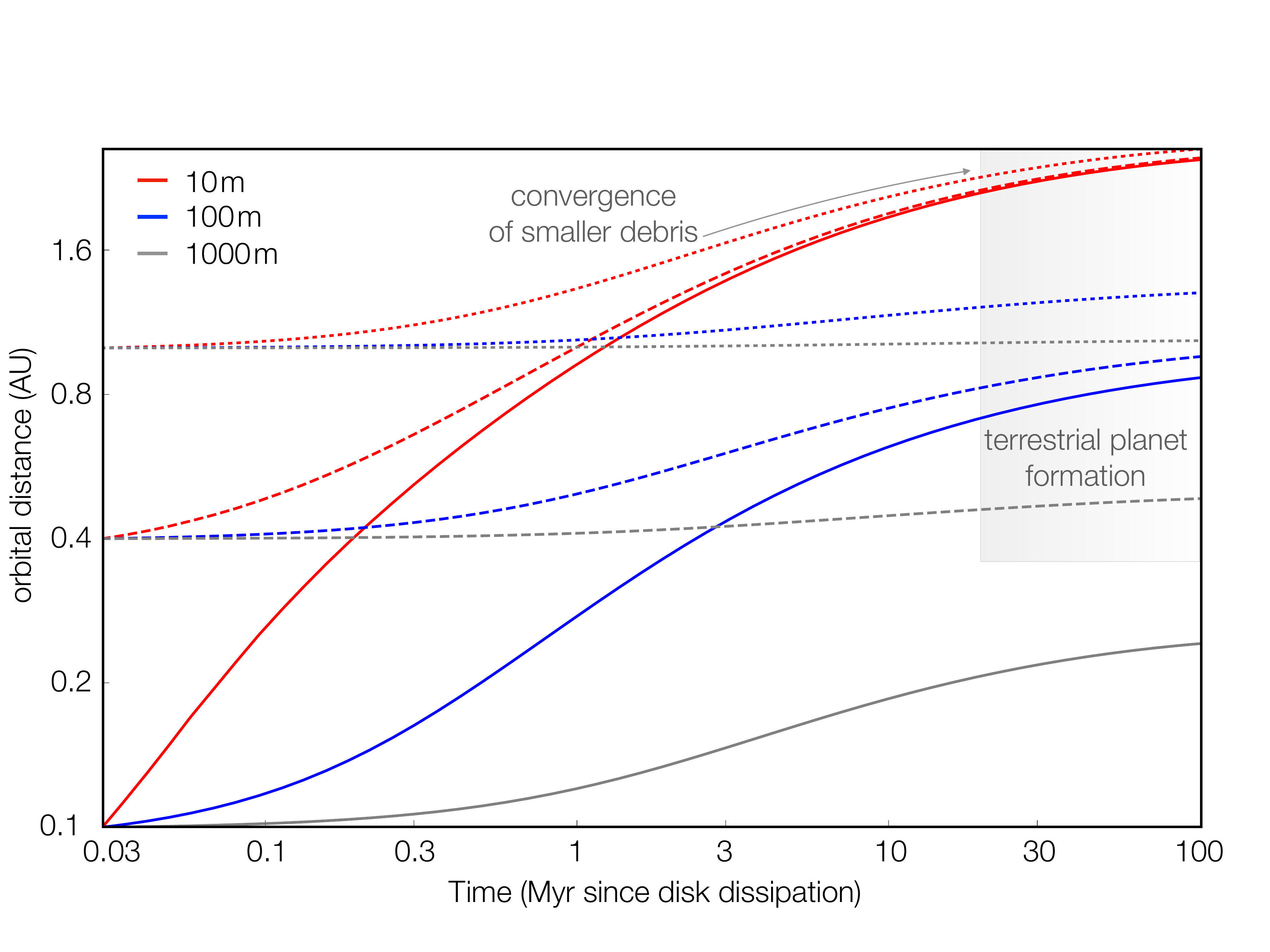}
\caption{Time evolution of 3 particle sizes (10\,m, 100\,m and 1000\,m) under the influence of the Solar wind, beginning at orbital distances of 0.1\,AU, 0.4\,AU and 1\,AU. Poynting-Robertson drag is included. We highlight the region and approximate time of terrestrial planet formation \citep{Kleine2009}, illustrating that this region may become enhanced by particles from the inner radii, depleting the inner Solar System in turn.}
\label{f4}
\end{figure}

We perform the integration until 100\,Myr, noting that the approximate formation time of Earth is much shorter, between 30-50\,Myr \citep{Kleine2009}. Accordingly, if the mass of material feeding Earth's formation was supplied largely from the inner Solar System, outward migration must occur within the first few 10s of millions of years. Figure~\ref{f4} illustrates the time evolution of simulated particles. For reference, the semi-major axis range and time occupied by Earth, Venus and Mercury's formation are shaded. Objects of 10\,m in size are transported well beyond even the orbit of Mars, with little dependence upon their initial locations. Objects of 100\,m size converge in their orbital locations closer to 1\,AU, and km-sized objects undergo a small degree of outward migration. 

Accordingly, we may conclude that during the early phases of planet formation, particles of 100\,meter radii and smaller are expelled by the Solar wind beyond Mercury's present orbit, even if once present at 0.1\,AU. Next, it is important to discuss how much material may have existed within 100\,m and smaller size classes in the Solar System. If it is significantly smaller than Mercury's mass, then the effect of the Solar Wind in polluting the outer Solar System is minimal, but if it rivals Earth's mass, the bulk of terrestrial planet material may have originated within the hottest, inner parts of the Solar System. In the next section, we show that the latter option is the more likely situation. 

  \section{Discussion \& Conclusions}
\subsection{Mass transported}
If the terrestrial planet-forming region was supplied by objects below $\sim100$\,m in size, it is important to consider the mass of material this amounts to, relative to the mass currently present in the planets themselves. The mass swept up by Jupiter has been proposed to reach 10--20\,M$_\oplus$ \citep{Batygin2015b}. This value is similar to the mass derived from integrating the solid component of the ``minimum mass extrasolar nebula" (MMEN; \citealt{Chiang2013}), the average surface density of solid material found in extrasolar planetary systems. Accordingly, we assume an order of magnitude value of 10\,M$_\oplus$ of material is subject to outward motion. 

Next, we consider what fraction of this available mass resides in size classes that are subject to the Solar wind. This calculation is more uncertain, owing to unknown conditions within the early Solar System. Jupiter's inward migration likely set off a collisional cascade that ground large objects down to smaller particles \citep{Wyatt2008,Batygin2015b}. Collisional cascades typically produce a distribution of particles where the number in any size class may be represented as a power-law. Given a power-law slope, the fraction of particles smaller than 100\,meters depends upon the largest sizes that exist in the population of debris.

We suppose that the number of particles $dN$ between sizes of $s$ and $s+ds$ is written \citep{Wyatt2008,Hughes2018}
\begin{align}
dN=g(s)ds\propto\,s^{-q}ds,
\end{align}
between a largest size $s_{\textrm{up}}$ and a smallest size $s_{\textrm{down}}$. If the Solar wind affects particles smaller than size $s_{\textrm{w}}$, the fraction of the mass contained within sizes $s_{\textrm{down}}$ and $s_{\textrm{w}}$ may be written (assuming independence between density and particle size)
\begin{align}
f(s_{\textrm{up}},s_{\textrm{w}})=\frac{s_{\textrm{w}}^{4-q}-s_{\textrm{down}}^{4-q}}{s_{\textrm{up}}^{4-q}-s_{\textrm{down}}^{4-q}}.
\end{align}
The choice of $s_{\textrm{down}}$ is not important provided it is small compared to both $s_{\textrm{w}}$ and $s_{\textrm{up}}$, and so we fix $s_{\textrm{down}}=10^{-5}\,$m for the sake of definiteness. We make the assumption that particle masses scale with $s^3$ and adopt a value $q=3.5$ \citep{Hughes2018}.

\begin{figure}[h!]
\centering
\includegraphics[trim=0cm 0cm 0cm 0cm, clip=true,width=1\columnwidth]{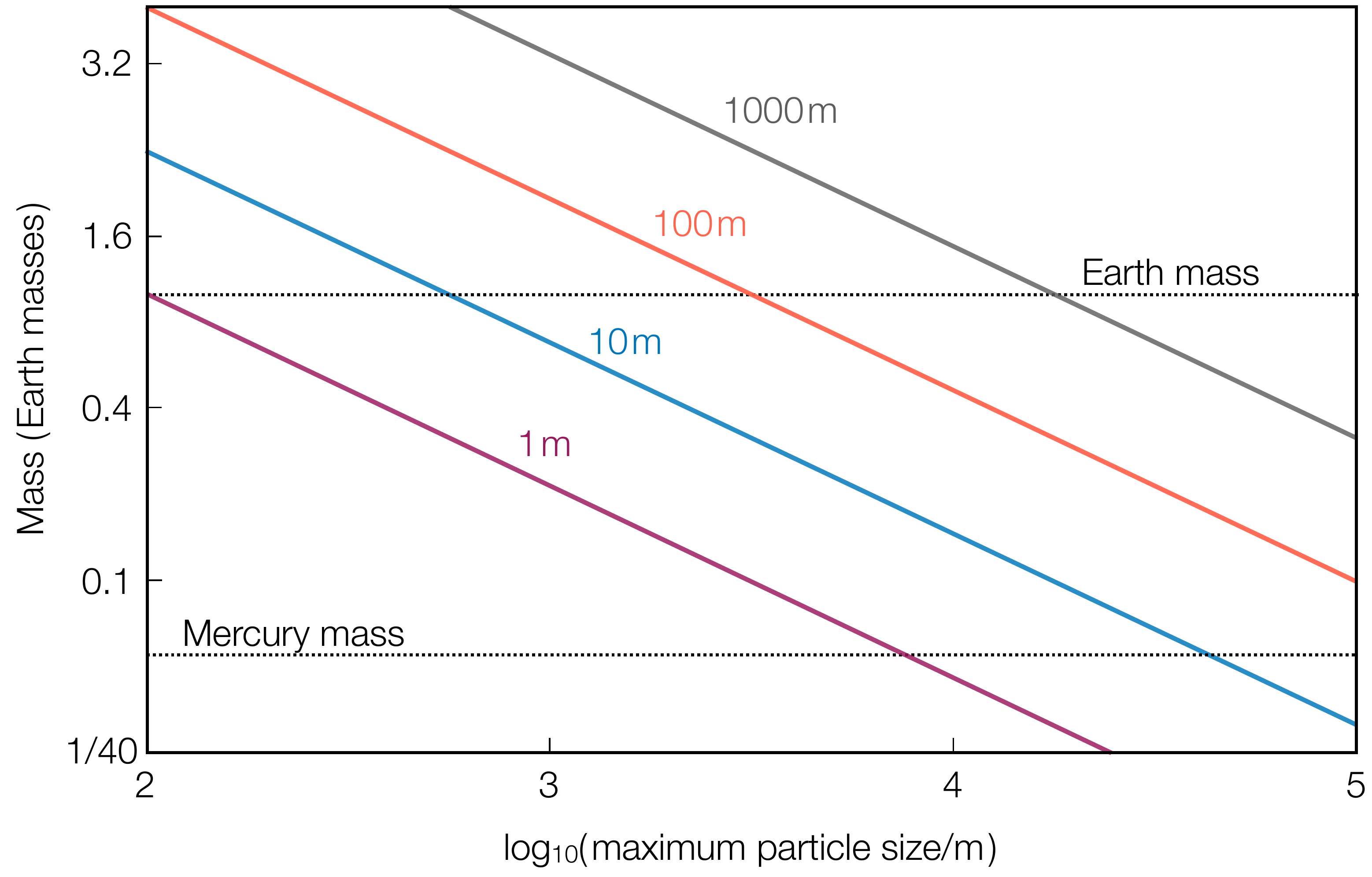}
\caption{The mass contained in particles smaller than 1m, 10m, 100m and 1000m as a function of the largest member of the collisional debris. Masses between Mercury and Earth are reasonable to be excavated, given the 10\,M$_{\oplus}$ hypothesized to have been swept up during Jupiter's inward migration \citep{Batygin2015b}.}
\label{f5}
\end{figure}

In Figure~\ref{f5}, we depict the amount of mass, scaled to Earth's mass, contained within particles smaller than 1\,m, 10\,m, 100\,m and 1\,km, as a function of the largest particles present in the debris. If the collisional cascade disrupts objects larger than around 100\,km \citep{Batygin2015b}, the Solar wind is able to expel a mass exceeding Mercury in 100\,m and smaller-sized objects, which we showed above to be susceptible to the Solar wind. More optimistic estimates, such as 1\,km-sized objects within a population of 10km-sized bodies may yield over an Earth's mass of material. 

Given nominal parameters, the terrestrial planet region may have been polluted by material from inside of Mercury's orbit totalling a mass that rivals that of the terrestrial planets themselves. The picture considered in this work rests on the assumption that Jupiter inhibited planet formation prior to disk-dispersal \citep{Batygin2015b} -- if instead the debris was consolidated into planets earlier, as in numerous extrasolar planetary systems, the Solar wind would not be sufficient to influence their orbits \citep{Lovelace2008}. Accordingly, the dichotomous mass-density of material within 0.4\,AU of the host star in extrasolar planetary systems, and our Solar System, naturally arises as a combination of disruption from a outer giant planet, combined with ram pressure from the early stellar wind \citep{Chiang2013} 

As an additional note, we mention that stellar contraction was omitted in our modeling. In reality Sun-like stars contract on 10s of millions of year timescales \citep{Shu1987}, weakening the solar wind-induced supply of planetary debris from the inner Solar System. Thus, the Solar wind may have ``regulated" the terrestrial planet-formation process, cutting off building blacks within 10s of millions of years. We leave this proposition for future investigation. 

\subsection{Pollution from the inner Solar System}

  If the outward migration of debris from the hot, innermost 0.4\,AU of the Solar System proposed here truly occurred, material residing on more distant orbits today should exhibit features consistent with a higher-temperature history. One potential example, is Mercury's high iron content could, which would arise from close-in sublimation of silicates, and not iron \citep{Kama2009}. Additionally, only a small fraction of the innermost debris needs to be transported as far as the asteroid belt to significantly affect the composition of its material. The oldest objects in the Solar System, the Calcium-Aluminium inclusions exhibit evidence for a high-temperature formation environment \citep{MacPherson2005}, and the presence of crystalline silicates within numerous chondritic meteorites is also suggestive of a high-temperature history \citep{Wooden2005}. 
  
  The Earth itself exhibits numerous compositional mysteries, including a relatively low carbon content \citep{Lee2010}, and Ruthenium isotopes more consistent with objects interior to its orbit, than exterior \citep{Fischer2017}. The Solar wind-driven outward migration of debris may explain some of these peculiar features within the colder reaches of the current Solar System. Morever, the influence of young stellar winds may be directly detectable in the form of leading magnetic tails upon close-in exoplanets. Indeed, such a leading tail has already been discovered \citep{Sanchis2015}, though the later age of the host star suggests a different cause in this specific case. 
  
   As new exoplanetary candidates are discovered, the opportunities to seek signatures of strong stellar winds will continue to grow, providing a key observational insight into the connection between the forces sculpting the inner regions of our own Solar System, and of systems around other stars.

\begin{acknowledgements}
 This research was funded by a 51 Pegasi b Postdoctoral Fellowship, of the Heising-Simons Foundation. We thank Konstantin Batygin, Greg Laughlin, Woodward Fischer and the referee for useful insights.
\end{acknowledgements}

%\bibliographystyle{apa}
%\bibliography{Wind_References} 

\end{document}